\documentclass[aps,nofootinbib,prd,preprint,preprintnumbers]{revtex4}

\pdfoutput=1
\usepackage{graphicx}

\usepackage{amsmath}
\usepackage{amssymb}

\usepackage{hyperref}
\usepackage{color}

\newcommand{\feyn}[1]{#1\kern-0.45em/}
\newcommand{\feynn}[1]{#1\kern-0.65em/}
\newcommand{\ket}[1]{|#1\rangle}

\newcommand{\hyp}[1]{(\hyperref[#1] {\ref{#1}})}
\newcommand{\dm}[1]{\Delta m^2_{#1}}

\renewcommand\Re{\operatorname{Re}}
\renewcommand\Im{\operatorname{Im}}

\newcommand{\Cinvestav}{Departamento de F\'{\i}sica, Centro de
  Investigaci{\'o}n y de Estudios Avanzados del IPN\\ Apdo. Postal
  14-740 07000 Mexico, DF, Mexico}
  
\begin{document}

\title{
Constant matter neutrino oscillations in a  parametrization-free formulation. \\
}

\author{L. J. Flores~$^1$}\email{jflores@fis.cinvestav.mx}
\author{O. G. Miranda~$^1$}\email{omr@fis.cinvestav.mx}
\affiliation{$^1$~\Cinvestav}

\begin{abstract}
Neutrino oscillations are now a well-stablished and deeply studied
phenomena. Their mixing parameters, except for the CP phase, are
measured with good accuracy. The three-neutrino oscillation picture in
matter is currently of great interest due to the different
long-baseline neutrino experiments that are already running or under
construction.  In this work, we reanalyze the exact expression for the
neutrino probabilities (in a constant density medium) and introduce an
approximate formula.  Our results are shown in a formulation that is
independent of the parametrization and could be useful for unitary
tests of the leptonic mixing matrix. We illustrate how the
approximation, besides being simple, can reproduce the neutrino
probabilities with good accuracy.
\end{abstract}

\maketitle
\section{Introduction}
Neutrino oscillations are a well-stablished phenomenon, with
parameters that have been measured with great
accuracy~\cite{Forero:2014bxa}, except for the CP-violating phase
which is expected to be precisely determined in the new generation of
long-baseline neutrino experiments.  Despite this success, all the
analysis has been done in a particular parametrization~\cite{PDG}. In
the case of neutrino physics, there are other parametrizations that
could be
interesting~\cite{Schechter:1980gr,Rodejohann:2011vc,Escrihuela:2015wra},
taking into account the possibility of accounting for more neutrino
families as well as the richness of the possible neutrino Majorana
nature.  Moreover, for a unitarity test~\cite{Parke:2015goa}, it may
be more interesting to analyze the experimental data in a
parameter-independent way, studying the values of the matrix
entrances. It is already known that, for three families of Dirac
fermions, the mixing matrix will have four independent
elements~\cite{Bjorken:1987tr,Dunietz:1987yt,Branco:1987mj,Wagner:1998mc,Giunti:2007ry}.
If neutrinos have Majorana a nature, or if there are more neutrino
families, the situation will be more complex~\cite{Valle:2015pba}, but
a test of the ``standard'' picture through unitarity can be considered
as a first step into the search for new physics.
On the other hand, an important part of neutrino data comes from
matter effects~\cite{Wolfenstein:1977ue}, making the analysis more
complicated due to the need for computing numerical solutions to the
neutrino evolution equation, for the case of varying density profiles,
or for using approximate formulas for neutrino probabilities in
constant density
environments~\cite{Barger:1980tf,Zaglauer:1988gz,Cervera:2000kp,Freund:2001pn,Akhmedov:2004ny,Minakata:2015gra}.

In this work, we first discuss the exact formula for neutrino
oscillations in a constant density environment. We show our results in
terms of the entrances of the leptonic mixing matrix and, therefore,
they are independent of the parametrization. Afterwards, we introduce
a new approximation that, besides being simple, can be formulated in
terms of a series expansions and, therefore, can be computed with a
level of accuracy according to the phenomenological needs of the given
problem.
The expressions found here could be useful in analyzing the neutrino
data either in the standard parametrization or in other contexts, such
as unitary test in the neutrino sector.

\section{The exact case}

The leptonic mixing matrix relates the mass and flavor states through
the relation $\nu_\alpha = \sum_i U_{\alpha i}\nu_i$, where the matrix
$U_{\alpha i}$ could be parametrized, for instance, in the usual
convention adopted by the Particle Data Group (PDG)~\cite{PDG}. The
evolution equation in vacuum will be given by
\begin{equation}\label{2.5}
	i\frac{d}{dt}\nu_j=\frac{m_j^2}{2E}\nu_j
\end{equation} 
That leads, through a well-known procedure, to the usual expression for 
neutrino probabilities in vacuum
\begin{eqnarray}\label{2.xx}
	P_{\nu_\alpha \rightarrow \nu_\beta}=
\delta_{\alpha \beta}-4\displaystyle\sum^{n}_{\ell >j} \Re\left[ U_{\alpha \ell}^{*}U_{\beta \ell}U_{\alpha j}U_{\beta j}^{*} \right] \sin^2\left( \frac{\Delta m^2_{\ell j}L}{4E}\right)& \nonumber \\ 
+2\displaystyle\sum^{n}_{\ell >j} \Im\left[ U_{\alpha \ell}^{*}U_{\beta \ell}U_{\alpha j}U_{\beta j}^{*} \right] \sin \left( \frac{\Delta m^2_{\ell j}L}{2E}\right).&
\end{eqnarray}
Notice however, that in the case when the extra neutrino states are heavy, they do not participate in the oscillation and the sum is cut up to three: 
\begin{eqnarray}\label{2.x1}
	P_{\nu_\alpha \rightarrow \nu_\beta}=
\sum^3_{\ell ,j} U^*_{\alpha \ell}U_{\beta \ell}U_{\alpha j}U^*_{\beta j} - 
4\displaystyle\sum^{3}_{\ell >j} \Re\left[ U_{\alpha \ell}^{*}U_{\beta \ell}U_{\alpha j}U_{\beta j}^{*} \right] \sin^2\left( \frac{\Delta m^2_{\ell j}L}{4E}\right)& \nonumber \\ 
+2\displaystyle\sum^{3}_{\ell >j} \Im\left[ U_{\alpha \ell}^{*}U_{\beta \ell}U_{\alpha j}U_{\beta j}^{*} \right] \sin \left( \frac{\Delta m^2_{\ell j}L}{2E}\right),&
\end{eqnarray}
and the $\delta_{\alpha\beta}$ appearing in Eq.~(\ref{2.xx}) is
substituted by the well-known zero distance effect.

If we would like to consider matter effects, we need to add the 
charged current potential due to electrons that, again in the mass basis:
\begin{equation} \label{2.8}
	i\frac{d}{dt}\nu_j=\frac{1}{2E}\left( m_j^2 \nu_j + \displaystyle\sum_k AU_{e j}^* U_{ek}\nu_k\right),
\end{equation}
\noindent with $A=2EV_{CC}$.  
It is known that in this case we can express the probability in a similar form 
\begin{eqnarray}\label{2.22}
	P_{\nu_\alpha \rightarrow \nu_\beta}=\delta_{\alpha \beta}-4\displaystyle\sum^{n}_{\ell >j} \Re\left[ V_{\alpha \ell}^{*}V_{\beta \ell}V_{\alpha j}V_{\beta j}^{*} \right] \sin^2\left( \frac{\Delta M^2_{\ell j}L}{4E}\right)& \nonumber \\ 
+2\displaystyle\sum^{n}_{\ell >j} \Im\left[ V_{\alpha \ell}^{*}V_{\beta \ell}V_{\alpha j}V_{\beta j}^{*} \right] \sin \left( \frac{\Delta M^2_{\ell j}L}{2E}\right).&
\end{eqnarray}
by defining 
\begin{equation} \label{2.20}
	V=UW^T,
\end{equation}
where $W$ is an unitary matrix. We can also find the correspoding
expression for Eq.~(\ref{2.x1}) in the presence of matter:
\begin{eqnarray}\label{2.22a}
	P_{\nu_\alpha \rightarrow \nu_\beta}=
\sum^3_{\ell ,j} U^*_{\alpha \ell}U_{\beta \ell}U_{\alpha j}U^*_{\beta j} -
4\displaystyle\sum^{3}_{\ell >j} \Re\left[ V_{\alpha \ell}^{*}V_{\beta \ell}V_{\alpha j}V_{\beta j}^{*} \right] \sin^2\left( \frac{\Delta M^2_{\ell j}L}{4E}\right)& \nonumber \\ 
+2\displaystyle\sum^{3}_{\ell >j} \Im\left[ V_{\alpha \ell}^{*}V_{\beta \ell}V_{\alpha j}V_{\beta j}^{*} \right] \sin \left( \frac{\Delta M^2_{\ell j}L}{2E}\right) , &
\end{eqnarray}
where, as expected, the zero distance term remains unchanged, thanks to the
unitarity of the $W$ matrix.

To find the expressions for the matrix $V$, we follow the procedure
described in Ref.~\cite{Zaglauer:1988gz}.  We will arrive to the same
expressions, except that we maintain the matrix elements of $U_{\alpha
  i}$ in a parameter-independent form. Although the following
procedure is straightforward, it will will allow to see the
nonunitary case in a more transparent way. Even if we work in the
standard parametrization, the expression will be useful, as the
numerical computations will be slightly simplified by substituting the
parametrization at the end.

We start by noticing that  the term inside the parenthesis in the 
right-hand side of equation~(\ref{2.8}) defines a matrix
\begin{equation} \label{2.14}
	\left(
	\begin{array}{ccc}
 A|U_{e1}|^2 & A U_{e1}^* U_{e2} & A U_{e1}^* U_{e3} \\
A U_{e2}^* U_{e1} & \dm{21} + A|U_{e2}|^2 &  A U_{e2}^* U_{e3} \\
A U_{e3}^* U_{e1} & A U_{e3}^* U_{e2} & \dm{31} + A|U_{e3}|^2 \\
	\end{array}\right)
\end{equation}
 with a  characteristic polynomial given by 
\begin{eqnarray}\label{2.10}
	&\lambda'^3 -\dfrac{\lambda'^2}{2E}(\dm{21}+\dm{31}+A)+\dfrac{\lambda'}{4E^2}\left[\dm{31}\dm{21}\phantom{gggggggggggggggggggg}\right. \nonumber \\
	 &+\left. A(\dm{21}(1-|U_{e2}|^2)+\dm{31}(1-|U_{e3}|^2))\right]- \dfrac{1}{8E^3}\dm{21}\dm{31}A|U_{e1}|^2=0&
\end{eqnarray}

\noindent where we have subtracted a term $m_1^2$ from the main
diagonal in order to simplify the equation. We redefine the eigenvalues $\lambda'$ as $\lambda=2E\lambda'$, which implies that $\lambda_i=M_i^2$.

\noindent As is already known, for a polynomial of the form
\begin{equation}
\label{eq:polynomial}
	\lambda^3-\alpha \lambda^2 + \beta \lambda - \gamma=0,
\end{equation} 

\noindent the solutions for $\lambda$ real, are given by 
\begin{equation}
\label{eq:roots}
	\lambda_n=\frac{\alpha}{3}+\frac{2}{3}\sqrt{\alpha^2-3\beta}\cos\left[\frac{1}{3}\arccos\left( \frac{2\alpha^3-9\alpha\beta+27\gamma}{2\sqrt{(\alpha^2-3\beta)^3}}\right)+\frac{2n\pi}{3} \right], \quad n=0,1,2,
\end{equation}

\noindent that in our case imply
\begin{eqnarray} \label{2.12}
  \alpha &=& \dm{21}+\dm{31}+A (|U_{e1}|^2+ |U_{e2}|^2 + |U_{e3}|^2) 
  \nonumber \\
  \beta &=& \dm{31}\dm{21}+ A\dm{21}(|U_{e1}|^2 + |U_{e3}|^2)+
   A\dm{31}(|U_{e1}|^2+ |U_{e2}|^2) \nonumber \\
   \gamma &=& A\dm{21}\dm{31}|U_{e1}|^2 \\
   \eta &=& \cos\left[\frac{1}{3}\arccos\left( \frac{2\alpha^3-9\alpha\beta+27\gamma}{2\sqrt{(\alpha^2-3\beta)^3}}\right)\right] \nonumber .
\end{eqnarray}

\noindent This leads us to the three eigenvalue equations which we are
going to label as:
\begin{eqnarray}\label{2.13}
	M_1^2 \equiv  \lambda_1 &=&  \frac{\alpha}{3}-\frac{1}{3}\sqrt{\alpha^2 -3\beta}\eta -\frac{\sqrt{3}}{3}\sqrt{\alpha^2 -3\beta}\sqrt{1-\eta^2}, \nonumber \\
	M_2^2 \equiv \lambda_2 &=&  \frac{\alpha}{3}-\frac{1}{3}\sqrt{\alpha^2 -3\beta}\eta +\frac{\sqrt{3}}{3}\sqrt{\alpha^2 -3\beta}\sqrt{1-\eta^2},  \\
	 M_3^2 \equiv \lambda_3 &=&  \frac{\alpha}{3}+\frac{2}{3}\sqrt{\alpha^2 -3\beta}\eta .  \nonumber 
\end{eqnarray}

\noindent In order to construct the diagonalizing matrix we need the
corresponding eigenvectors that will be given by 
\begin{equation} \label{2.15}
	\ket{\lambda_1}= \frac{1}{C_1}\left(\begin{array}{c}
\Lambda_1 \\
AU_{e2}^*U_{e1}(M_1^2-\dm{31}) \\
AU_{e3}^*U_{e1}(M_1^2-\dm{21}) \\
	\end{array} \right) \quad \ket{\lambda_2}= \frac{1}{C_2}\left(\begin{array}{c}
A U_{e1}^*U_{e2}(M_2^2-\dm{31}) \\
\Lambda_2 \\
A U_{e3}^*U_{e2}M_2^2 \\
	\end{array} \right), 
\end{equation}
\begin{equation*}	
	 \ket{\lambda_3}= \frac{1}{C_3}\left(\begin{array}{c}
A U_{e1}^*U_{e3}(M_3^2-\dm{21}) \\
A U_{e2}^*U_{e3}M_3^2  \\
\Lambda_3\\
	\end{array} \right). 
\end{equation*}
Here, we define the normalization constants, $C_j$, as 
\begin{equation}\label{2.17}
C_j=\sqrt{\Lambda_j^2+A^2|U_{ej}|^2\displaystyle\sum_{i\neq j}|U_{ei}|^2(M_j^2-\dm{k1})^2},\quad \mbox{for}\quad k\neq i 
\end{equation}
and we also define
\begin{equation}\label{2.16}
	\Lambda_j= M_j^4- \displaystyle\sum_{i\neq j}\left[M_j^2\left( \dm{i1}+A|U_{ei}|^2\right)-A \dm{i1} |U_{ek}|^2-\frac{1}{2}\dm{i1} \dm{k1}\right], \quad \mbox{for} \quad k\neq i .
\end{equation}
Now we can write the explicit form of the matrix $W$, that in abbreviated form 
can be written as 
\begin{equation}\label{2.21}
(W^T)_{k j}= \frac{\Lambda_k}{C_k}\delta_{k j}
+(1-\delta_{k j})A \frac{U_{ek}U_{ej}^*\left(M_k^2-\sum_i[\dm{i1}\epsilon^2_{ijk}] \right)}{C_k}  
\end{equation}	
or, writing it explicitly, 
\begin{equation} \label{2.18}
	W= \left( \begin{array}{ccc}
\frac{\Lambda_1}{C_1} & \frac{A U_{e1}^*U_{e2}(M_2^2-\dm{31})}{C_2} & \frac{A U_{e1}^*U_{e3}(M_3^2-\dm{21})}{C_3}  \\
\frac{AU_{e2}^*U_{e1}(M_1^2-\dm{31})}{C_1} & \frac{\Lambda_2}{C_2} & \frac{A U_{e2}^*U_{e3}M_3^2}{C_3} \\
\frac{AU_{e3}^*U_{e1}(M_1^2-\dm{21})}{C_1} & \frac{A U_{e3}^*U_{e2}M_2^2}{C_2}  & \frac{\Lambda_3}{C_3}
	\end{array} \right).
\end{equation}
We have arrived to the explicit form of the diagonalizing matrix $W$, such that, 
\begin{equation}\label{2.19}
	W^{-1}H_MW= \frac{1}{2E}\left(\begin{array}{ccc}
M_1^2 & 0 & 0 \\
0 & M_2^2 & 0 \\
0 & 0 & M_3^2
	\end{array} \right).
\end{equation}
This matrix relates the mass states in vacuum with the matter ones in
the form $\ket{\nu '_M}=W\ket{\nu_M}$, where the primed vector refers
to the matter mass states.  It is easy to see that the vacuum case is
restored when $A=0$. With this relation we can find the oscillation
probabilities in matter as a function of the elements of the vacuum
rotation matrix, without the use of any parameterization and without
using the unitary relation.  Therefore, they could be useful to study
the unitarity of the mixing matrix, a topic that could be of interest
now that we are entering into a precision era in neutrino physics. As
we have already mentioned, this method is well
known~\cite{Zaglauer:1988gz}, although the treatment had been done in
a specific parametrization.

\section{An approximation}

Once we have discussed the exact solution for the constant density
matter case, we proceed to find an approximate formula for the
probabilities. In order to preserve the parametrization-free
structure, we look for an approximation for the cubic roots
$\lambda_i$ in Eq.~(\ref{eq:roots}). 

We start by noticing, from Eq.~(\ref{2.12}), that if $\Delta
m^2_{21}\to 0$, then $\gamma \to 0$ and the cubic equation
(\ref{eq:polynomial}) is reduced the quadratic case. If this is the
case, Eq.~(\ref{2.13}) will reduce to the two typical solutions for a
quadratic equation plus a third solution, given by $\lambda_1 = 0$. In
particular, we will have the expression for $\eta$:
\begin{equation}
\eta  
= \cos\left[\frac{1}{3}\arccos\left( \frac{2\alpha^3-9\alpha\beta}{2\sqrt{(\alpha^2-3\beta)^3}}\right)\right] .
\end{equation}
In this simple case it is easy to find that 
\begin{equation}
\eta 
= \cos\theta 
= \frac{-\frac{1}{2}\alpha}{\sqrt{\alpha^2-3\beta}} .
\end{equation}
Now we can consider that $\gamma$ is not zero, but it is ``small'', say
$\gamma<<\alpha\beta$. This seems a natural hypothesis since $\Delta
m^2_{21} << \Delta m^2_{31} $.  We can try to find the correction
$\varepsilon$ that fulfills both
\begin{equation}
\cos\theta 
= \frac{-\frac{1}{2}\alpha + \varepsilon}{\sqrt{\alpha^2-3\beta}}
\end{equation}
and 
\begin{equation}
\cos3\theta = 4\cos^3\theta - 3 \cos\theta \simeq  
\frac{2\alpha^3-9\alpha\beta+27\gamma}{2\sqrt{(\alpha^2-3\beta)^3}} . 
\end{equation}
If we work only up to first-order terms, below
$(\frac{\gamma}{\beta})^2$, it is easy to find that
\begin{equation}
\label{eq:firstorder}
\varepsilon = \frac{3 \gamma}{2\beta}
\end{equation}
fits both conditions. Therefore, the eigenvalues will be approximately given by 
\begin{eqnarray}\label{eq:eigenvalApprox}
	 M_1^2 \equiv \lambda_3 &\simeq&  \frac{2}{3}\varepsilon.  \nonumber \\
M_2^2 \equiv  \lambda_1 &\simeq&  
         \frac{1}{2}(\alpha-\frac{2}{3}\varepsilon) 
       -\frac{1}{2}\sqrt{(\alpha+\frac{2}{3}\varepsilon)^2 
         -4[\beta+(\frac{2}{3}\varepsilon)^2]},  \\
M_3^2 \equiv \lambda_2 &\simeq&  
         \frac{1}{2}(\alpha-\frac{2}{3}\varepsilon) 
       +\frac{1}{2}\sqrt{(\alpha+\frac{2}{3}\varepsilon)^2 
       -4[\beta+(\frac{2}{3}\varepsilon)^2]}, \nonumber
\end{eqnarray}
This seems to be a reasonable approximation that leads to the equation 
\begin{equation}
\label{eq:PolynomialEpsilon}
	\lambda^3-\alpha \lambda^2 + \beta \lambda 
        - (\frac{2}{3}\beta\varepsilon - \frac{4}{9} \alpha\varepsilon^2+
          \frac{8}{27}\varepsilon^3)=0 .
\end{equation} 
With the expression of $\varepsilon$ at first-order in $\frac{\gamma}{\beta}$, Eq.~(\ref{eq:firstorder}), 
we have 
\begin{equation}
\label{eq:PolynomialFirstOrder}
	\lambda^3-\alpha \lambda^2 + \beta \lambda 
        - (\gamma-\frac{\alpha \gamma^2}{\beta^2}+\frac{\gamma^3}{\beta^3})=0 .
\end{equation} 
We can go one step further and find the expression for $\varepsilon$ at second 
order in $\frac{\gamma}{\beta}$. In this case we propose that 
\begin{equation}
\label{eq:SecondOrder}
\varepsilon = \frac{3\gamma}{2\beta} + a_2(\frac{\gamma}{\beta})^2
\end{equation}
and demand that Eq.~(\ref{eq:PolynomialEpsilon}) reduces to the usual
cubic expression, Eq.~(\ref{eq:polynomial}), up to second-order terms.
This condition is fulfilled when $a_2 =
\frac{3}{2}\frac{\alpha}{\beta}$. Therefore, at second-order we have
\begin{equation}
\label{eq:EpsSecondOrder}
\varepsilon = \frac{3 \gamma}{2\beta} + \frac{3\alpha}{2\beta}\frac{\gamma^2}{\beta^2} .
\end{equation}
We can continue with this procedure and find recursively the
coefficients $a_k$ for any order of approximation that we would like
to have. That is, we can write $\varepsilon$ as an infinite polynomial
that, in principle, should give an exact solution. The polynomial
would have the form 
\begin{equation}
\label{eq:AnyOrder}
\varepsilon = \sum_{k=1}^\infty a_k(\alpha,\beta) [\frac{\gamma}{\beta}]^k 
\end{equation}
with
\begin{eqnarray}
\label{eq:coefficients}
a_1 &=& \frac{3}{2} \nonumber \\  
a_2 &=& \frac{3\alpha}{2\beta} \nonumber \\  
a_k(\alpha,\beta) &=& \frac{3}{2\beta}
             \left(\frac{4\alpha}{9} \sum_{\substack{i,j\\i+j=k}}a_ia_j 
              -\frac{8}{27} \sum_{\substack{i,j,l\\i+j+l=k}}a_ia_j a_l \right) ; 
  \,\,\, k>2
\end{eqnarray}

\begin{figure}
\centering
\includegraphics[width=0.4\linewidth]{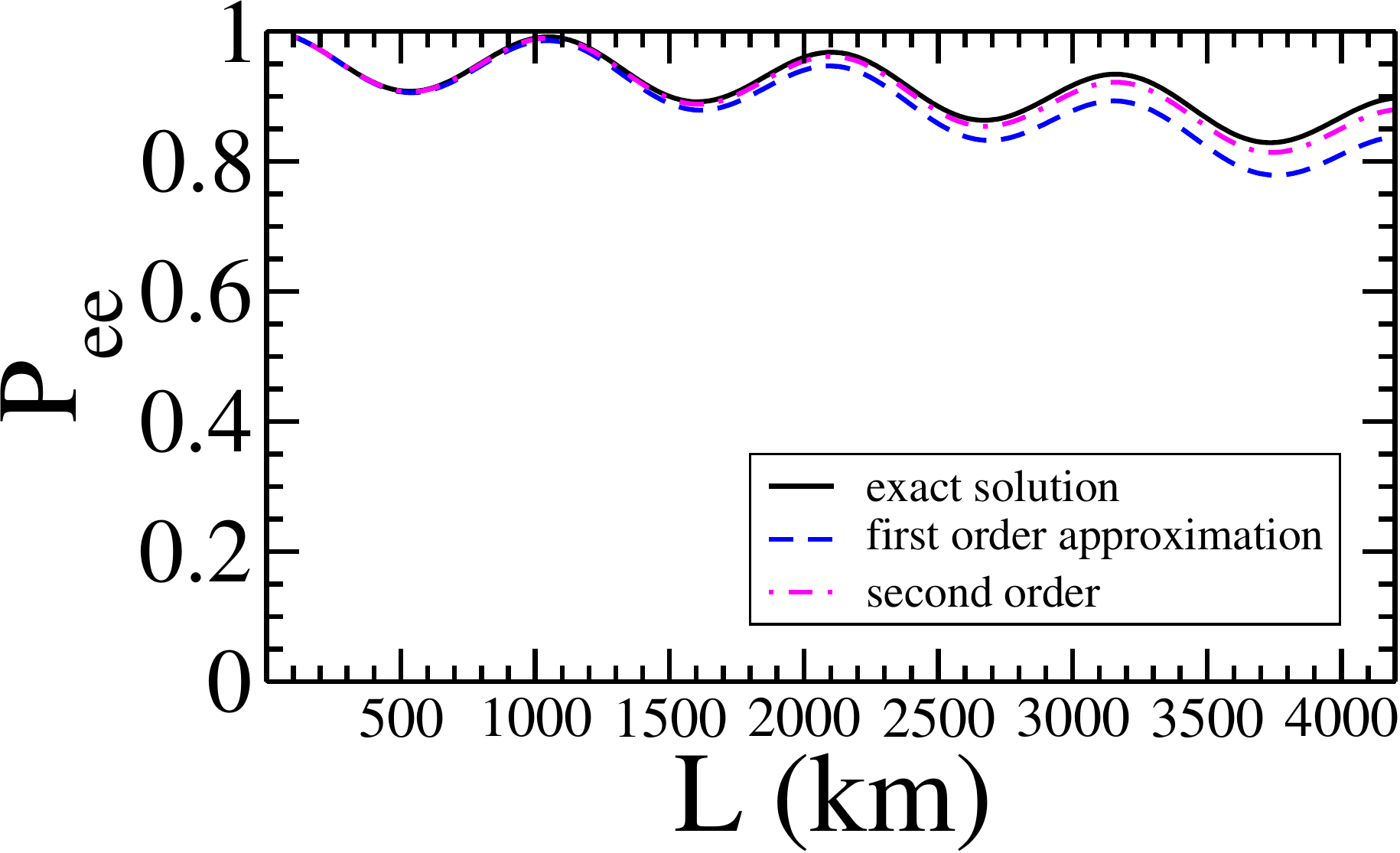}
\includegraphics[width=0.4\linewidth]{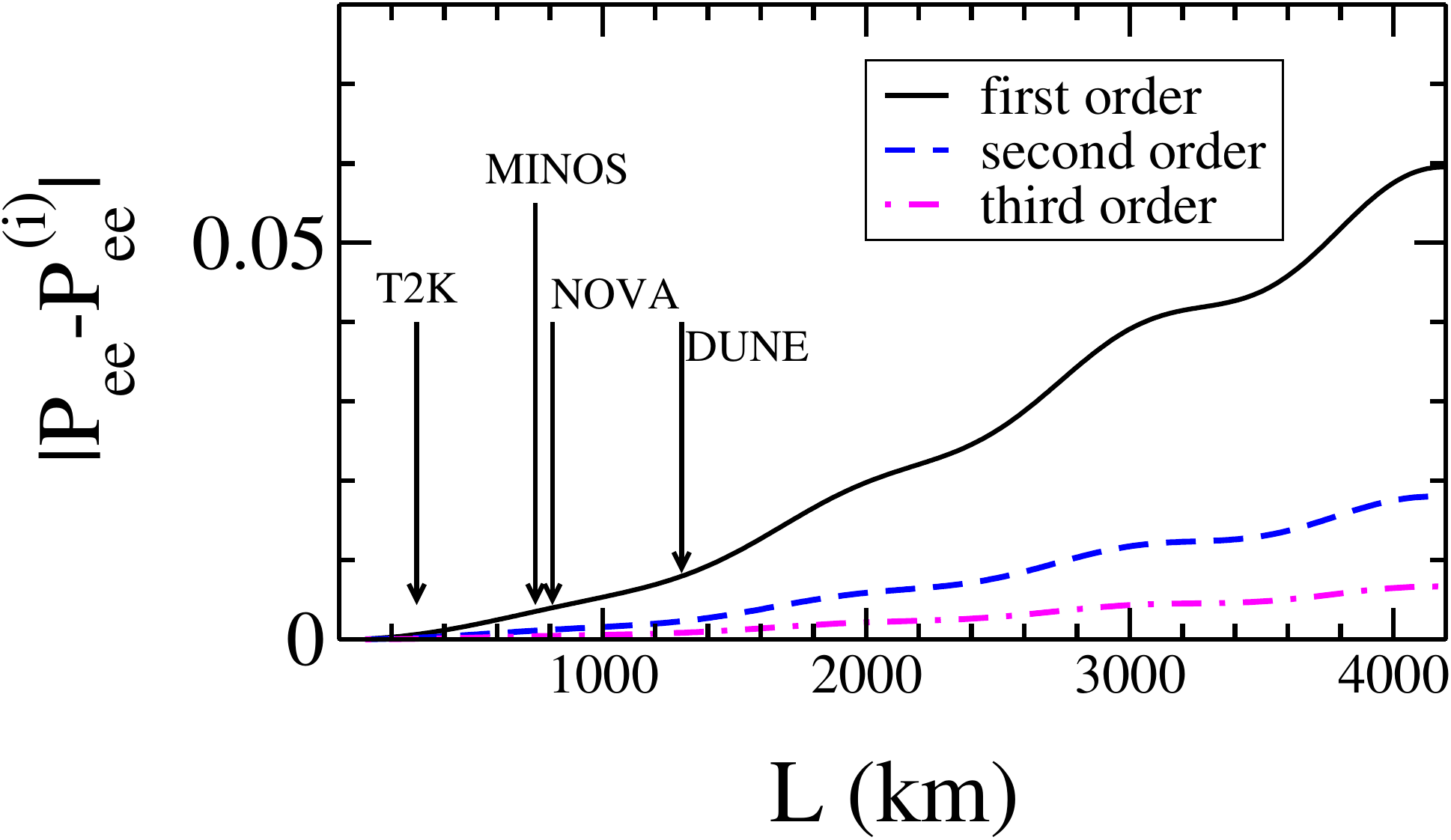}
\caption{Comparison of the exact electron neutrino survival
  probability in the three flavor case. The left panel shows the exact
  survival probability prediction for the central values of the mixing
  angles and mass squared differences and the approximated result for
  our approximation at order one and two.  
  In the right panel we show the absolute difference between the exact
  solution and the approximated prescription at first
  ($P_{ee}^{(1)}$), second ($P_{ee}^{(2)}$), and third order
  ($P_{ee}^{(3)}$).
  The baseline for different experiments and for the future
  DUNE experimental proposal is shown as a reference. The neutrino
  energy has been fixed to $E_\nu = 1$~GeV and the electron
  density has been taken to be $5.92\times 10^9$~eV$^3$. }
\label{fig:PEEabs}
\end{figure}

\begin{figure}
\centering
\includegraphics[width=0.4\linewidth]{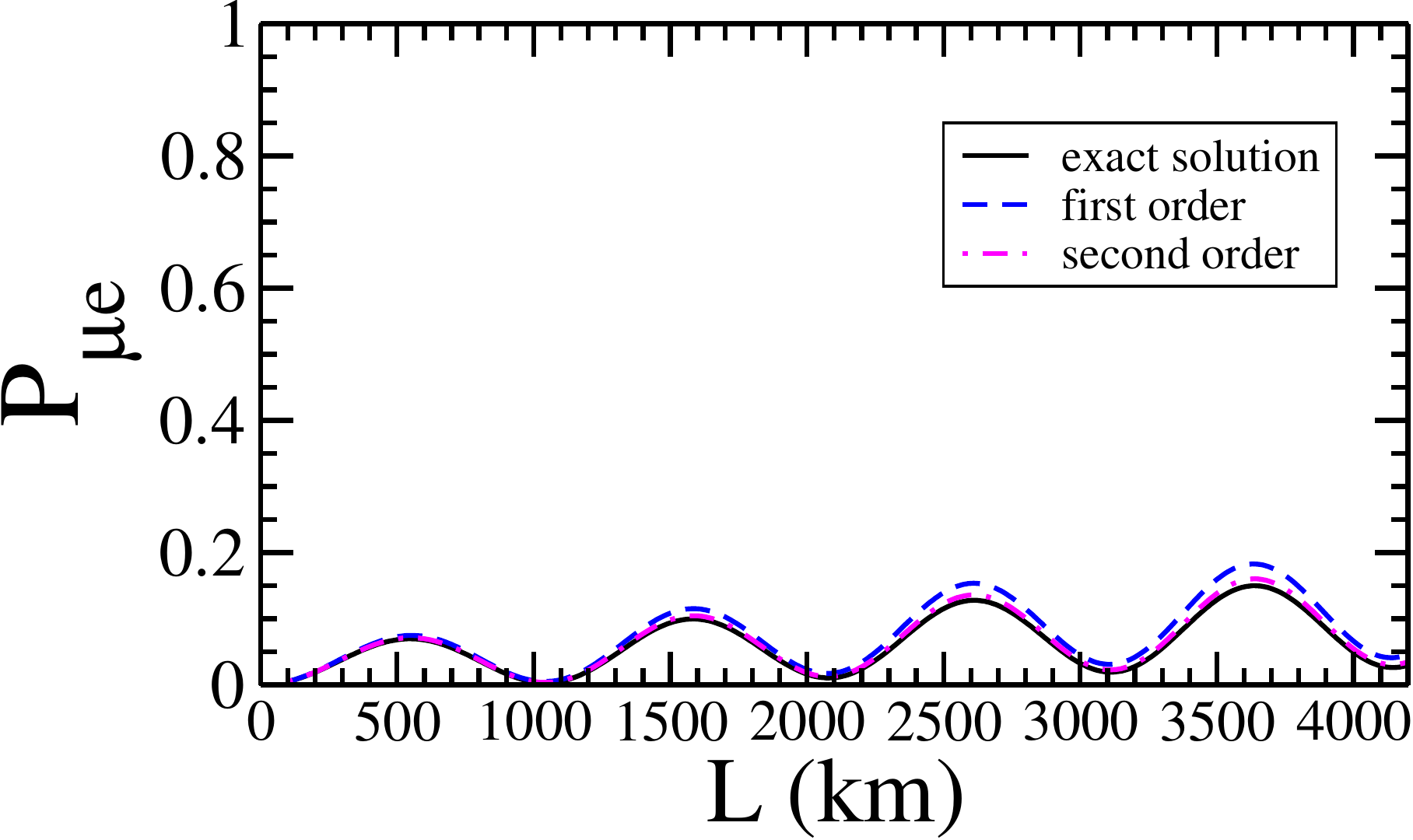}
\includegraphics[width=0.4\linewidth]{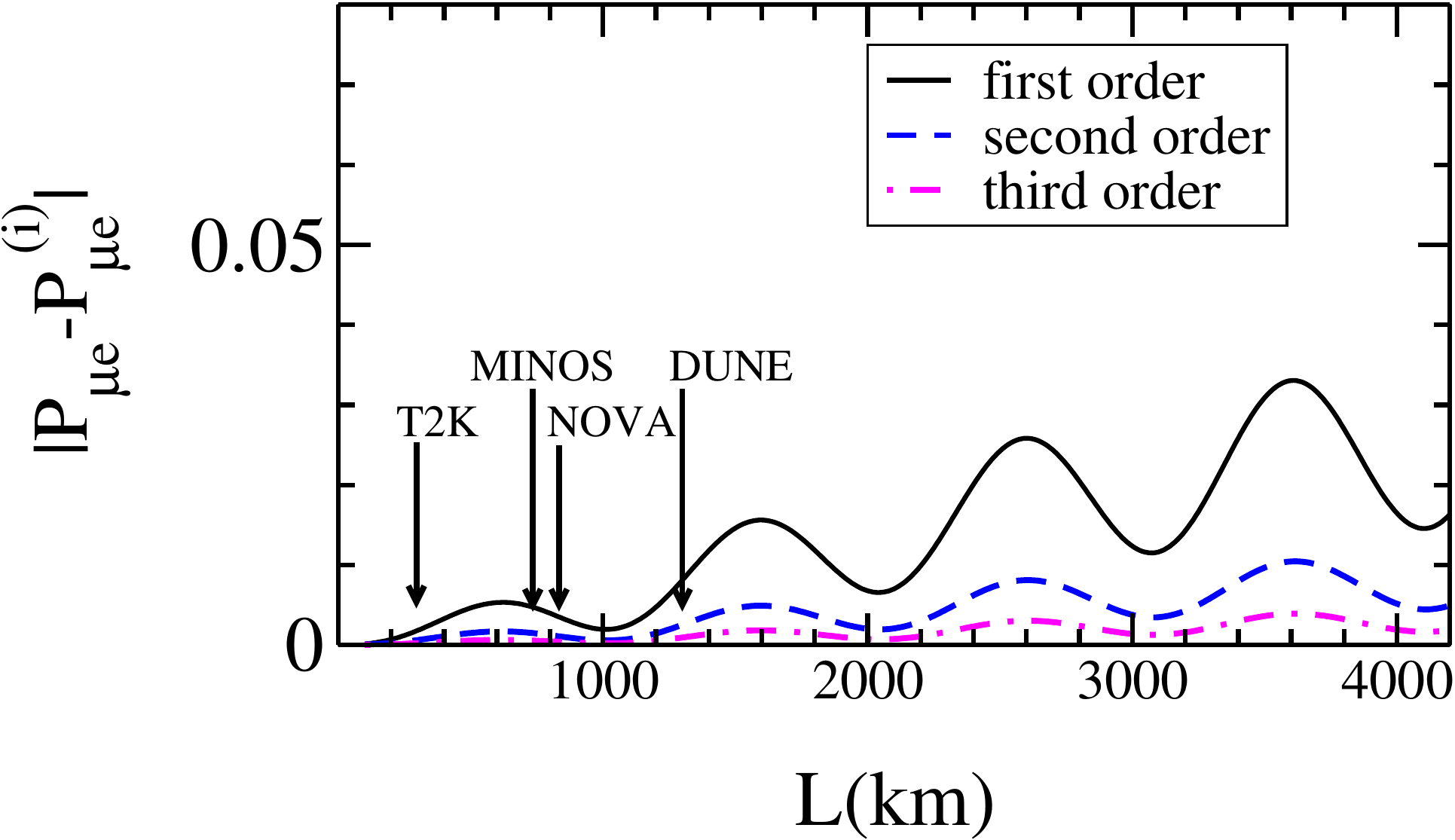}
\caption{Comparison of the exact muon to electron neutrino conversion 
  probability in the three flavor case. The left panel shows the exact
  survival probability prediction for the central values of the mixing
  angles and mass squared differences and the approximated result for
  our approximation at order one and two. 
  In the right panel we show the absolute difference between the exact
  solution and the approximated prescription at first
  ($P_{\mu e}^{(1)}$), second ($P_{\mu e}^{(2)}$), and third order
  ($P_{\mu e}^{(3)}$).
  The baseline for different experiments and for the future DUNE
  experimental proposal is shown as a reference. The neutrino energy
  has been fixed to $E_\nu = 1$~GeV and the electron density has been
  taken to be $5.92\times 10^9$~eV$^3$. }
\label{fig:PMUEabs}
\end{figure}

Once we have defined the approximation, we would like to know how well
it behaves with respect to the exact formula.  Although we have worked
out all the computation in a formulation that is independent of the
parametrization, we adopt now the standard PDG~\cite{PDG}
parametrization in order to substitute the current values for the
neutrino oscillation parameters. Therefore, in this case, we
explicitly adopt the unitary condition by making the following
substitutions in Eq.~(\ref{2.12})
\begin{eqnarray}
|U_{e1}|^2+ |U_{e2}|^2 + |U_{e3}|^2 &\equiv& 1 \nonumber \\
|U_{e1}|^2 + |U_{e3}|^2 &\equiv& 1 - |U_{e2}|^2 \\
|U_{e1}|^2 + |U_{e2}|^2 &\equiv& 1 - |U_{e3}|^2 \nonumber . 
\end{eqnarray}
Once we introduce the standard parametrization for the mixing matrix,
$U$, we adopt as central values of the mixing angles the ones reported
by Ref.~\cite{Forero:2014bxa} ($\sin^2\theta_{12}=0.320$ ,
$\sin^2\theta_{23}=0.613$, $\sin^2\theta_{13}=0.0246$) as well as the
corresponding squared mass differences ($\Delta m^2_{21}=7.62\times
10^{-5}$~eV$^2$, $\Delta m^2_{31}=2.55\times 10^{-3}$~eV$^2$). For the
value of the CP phase we have taken $\delta = 3\pi / 2$.
We have computed the survival probability $P_{ee}$ and the conversion
probability $P_{\mu e}$ and compared our approximated results with the
exact formulation, for a neutrino energy of $1$~GeV. The results are
shown in Fig.~(\ref{fig:PEEabs}) and Fig.~(\ref{fig:PMUEabs}) where we
have plotted these probabilities as functions of the baseline. From
these figures, it is possible to notice that the approximation works
reasonably well at first-order (especially for baselines below one
thousand kilometers) and has a great improvement when we consider
next-order approximations.

\begin{figure}
\centering
\includegraphics[width=0.6\linewidth]{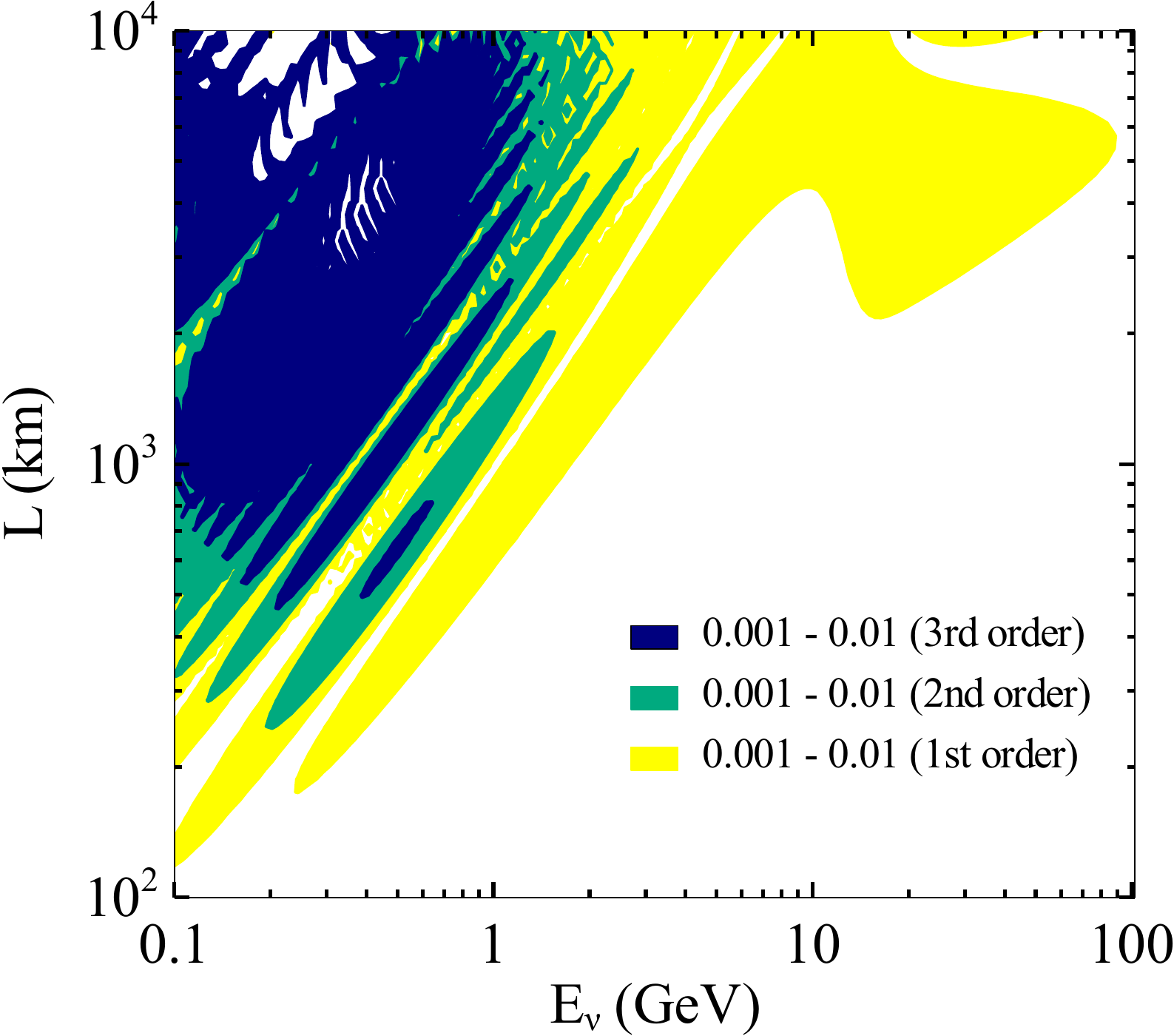}
\caption{Absolute difference between the exact conversion probability
  and the approximation discussed in this work. We have considered
  conversion from electron to muon neutrinos and computed the first,
  second, and third-order approximation for the oscillation parameters
  discussed in the text. We show the regions where the absolute
  difference lies in the range between $0.001$ and $0.01$. It is
  possible to see that the approximation works well for a wide range of
  values of distance and energy, especially at third-order.  }
\label{fig:comparison}
\end{figure}

\section{Conclusions}
In this work we have considered the case of three neutrino evolution in
a constant matter potential. We have first reviewed the exact
formulation and wrote the standard neutrino probabilities in a
parametrization-free scheme.  We have obtained an approximated formula
for this scenario that can be easily extended to the desired order of 
approximation, based on the coefficients for the eigenvalue problem, instead 
of considering specific oscillation parameters, such as $\Delta m^2_{21}$. 
This approximation can be used either for the
parametrization-free scenario (that could be useful in unitarity tests)
or in a particular parametrization such as the one adopted by the
PDG. We have shown that the formalism is simple and can be worked out
at any order of approximation, depending on the needs of the specific
problem.
The formalism could also be used for scenarios of physics beyond
the Standard Model such as the case of extra neutral heavy
leptons~\cite{Escrihuela:2015wra,deGouvea:2015euy}.

Finally, we can study the validity of the three orders of
approximation for different energies and baselines. In order to
compare with other results~\cite{Asano:2011nj}, we use the electron
neutrino conversion probability into muon neutrinos
and compute the absolute difference between our approximation and the
exact conversion formula.  Our results are summarized in
Fig.~(\ref{fig:comparison}) where we show the regions with an absolute
difference in the range $0.001-0.01$.  We use the oscillation
parameters already quoted above.  Comparing this result with the
approximation discussed in Ref.~\cite{Asano:2011nj} it is possible to
notice that our formula, at first-order, is not competitive in this
channel; however, for second and third-order, our approximation works
well, especially for energies at one GeV and above.


\section*{Acknowledgements}
This work has been supported by the CONACyT Grant No. 166639 (Mexico).

\end{document}